\documentstyle[epsf,multicol,aps]{revtex}

\def\l{\langle}
\def\r{\rangle}

\begin{document}
\draft
\title{
Probability-Changing Cluster Algorithm for Potts Models
}

\author{Yusuke Tomita\cite{tomita} and Yutaka Okabe\cite{okabe}}
\address{
Department of Physics, Tokyo Metropolitan University,
Hachioji, Tokyo 192-0397, Japan
}

\date{Received \today}

\maketitle

\begin{abstract}
We propose a new effective cluster algorithm of tuning 
the critical point automatically, which is 
an extended version of Swendsen-Wang algorithm. 
We change the probability of connecting spins of 
the same type, $p = 1 - e^{- J/ k_BT}$, in the process of 
the Monte Carlo spin update. 
Since we approach the canonical ensemble asymptotically, we can use 
the finite-size scaling analysis for physical quantities 
near the critical point.  Simulating the two-dimensional 
Potts models to demonstrate the validity of the algorithm, 
we have obtained the critical temperatures and critical exponents 
which are consistent with the exact values; the comparison has 
been made with the invaded cluster algorithm. 
\end{abstract}

\pacs{PACS numbers: 05.50.+q, 64.60.Ak, 75.10.Hk}

\begin{multicols}{2}
\narrowtext

Cluster algorithms \cite{sw87,wolff88} have been 
successfully used to overcome slow dynamics in 
the Monte Carlo simulation.  Swendsen and Wang (SW)
\cite{sw87} applied the Kasteleyn-Fortuin (KF) \cite{kf69} 
representation to identify clusters of spins.  
The problem of the thermal phase transition is mapped onto 
the geometric percolation problem in the cluster formalism 
\cite{kf69,ck80,hu84}.  Quite recently, based on the 
cluster formalism, the multiple-percolating clusters 
of the Ising system with large aspect ratio have been 
studied \cite{toh99}.

Machta {\it et al.} \cite{machta95} proposed another 
type of cluster algorithm, which is called the invaded cluster 
(IC) algorithm; this algorithm samples the critical point of a spin system 
without {\it a priori} knowledge of the critical temperature. 
It is in contrast with the usual procedure that one makes 
simulations for various parameters to determine the 
critical point.  The IC algorithm has been 
shown to be efficient in studying various physical quantities 
in the critical region, but the ensemble is not necessarily clear. 
Moreover, it has a problem of ``bottlenecks", 
which causes the broad tail in the distribution of 
the fraction of the accepted satisfied bonds \cite{machta95}. 

In this Letter, extending the SW algorithm \cite{sw87}, 
we propose a new algorithm of tuning the critical point 
automatically.  The basic idea of our algorithm is that 
we change the probability of connecting spins of 
the same type, $p = 1 - e^{- J/ k_BT}$, in the process of 
the Monte Carlo spin update, where $J$ is the exchange coupling. 
We decrease or increase $p$ depending on the observation 
whether the KF clusters are percolating or not percolating.
This simple negative feed-back mechanism together with 
the finite-size scaling (FSS) property of the existence 
probability (also called the crossing probability) $E_p$, 
the probability that the system percolates, 
leads to the determination of the critical point. 
Since our ensemble is asymptotically canonical 
as $\Delta p$, the amount of the change of $p$, becomes 0, 
the distribution functions of physical quantities obey the FSS; 
as a result, we can determine critical exponents 
using the FSS analysis.

Let us explain the procedure for our probability-changing cluster 
(PCC) algorithm in detail.  As an example, we consider 
the ferromagnetic $q$-state Potts model \cite{Wu82} 
whose Hamiltonian is given by
\begin{equation}
  {\cal H} = -J \sum_{<i,j>} (\delta_{\sigma_i,\sigma_j}-1), \quad 
  \sigma_i = 1, 2, \cdots, q,
\label{Hamiltonian}
\end{equation}
and for $q$ = 2 this corresponds to an Ising model.
The procedure of Monte Carlo spin update is as follows:
\begin{enumerate}
\item
Start from some spin configuration and some value of $p$. 
\item
Construct the KF clusters using the probability $p$, and check 
whether the system is percolating or not. 
Update spins following the same rule as the SW algorithm, 
that is, flip all the spins on any KF cluster to one of 
$q$ states.
\item
If the system is percolating (not percolating) in the previous test, 
decrease (increase) $p$ by $\Delta p \ (>0)$.
\item
Back to the process 2.
\end{enumerate}

After repeating the above processes, the distribution of $p$ 
for our Monte Carlo samples approaches the Gaussian distribution 
of which mean value is $p_c(L)$;  $p_c(L)$ is the probability 
of connecting spins, such that the existence probability 
$E_p$ becomes 1/2.  The width of the distribution depends on 
the choice of $\Delta p$ in the process 2, and becomes 0 
in the limit of $\Delta p \rightarrow 0$.  
We should note that $p_c(L)$ depends on the system size $L$, 
and $E_p$ follows the FSS near the critical point, 
\begin{equation}
  E_p(p,L) \sim X(t L^{1/\nu}), \quad t=(p_c-p)/p_c ,
\label{scale}
\end{equation}
where $p_c$ is the critical value of $p$ for the infinite system 
($L \rightarrow \infty$), and $\nu$ is 
the correlation-length critical exponent. 
(As for the FSS of $E_p$, see Ref.~\cite{hlc95}, for example.) 
We can estimate $p_c$ from the size dependence of $p_c(L)$ 
using Eq.~(\ref{scale}) and, in turn, estimate $T_c$ 
through the relation $p_c = 1 - e^{- J/ k_BT_c}$. 
We have chosen the value of $E_p$ which gives $p_c(L)$ as 1/2 
because it is the simplest.
We may modify the update process such that this value is 
different from 1/2. 

A comment should be made here on the choice 
of criterion to determine percolating.  
Machta {\it et al.} \cite{machta95} used both the extension 
rule and the topological rule for their stopping condition.
The former rule is that some cluster has maximum extent $L$ 
in at least one of the $d$ directions in $d$-dimensional systems.
The latter rule is that some cluster winds around the system 
in at least one of the $d$ directions.  We can use any rule 
to determine percolating, but FSS functions for physical quantities, 
therefore $p_c(L)$, depend on the rule.  

There is one free parameter in our algorithm; we may choose 
the difference $\Delta p$ in the process 3. 
In the limit of small $\Delta p$ we obtain the canonical 
ensemble, but it takes a long time to equilibrate
for small $\Delta p$.  Practically, we may start with 
rather large $\Delta p$, and switch to smaller $\Delta p$ 
with monitoring the trail of the values of $p$.  
Small steps of preparation are enough for equilibration.

In order to show the validity of the present method, we have 
made simulations for the $2d$ ferromagnetic 2-state Potts model 
(Ising model) and 3-state Potts model.  We have treated 
the systems with linear sizes $L$ = 64, 128, 256, and 512. 
We start with $\Delta p=0.01$, and gradually decrease $\Delta p$ 
to the final value.  We have chosen this final value of $\Delta p$ 
as $1/(20 \times L^2)$; the steps for preparation are 10,000 
for the largest size ($L=512$).  After reaching the final small 
value of $\Delta p$, we have taken 100,000 (200,000) Monte Carlo 
samples in the case of $q=2$ ($q=3$) with keeping $\Delta p$ 
as constant.  From each bond configuration, we have made 
5 (10) spin configurations in order to get better statistics 
for magnetization in the case of $q=2$ ($q=3$). 
We have performed several runs for each size, 
and have checked the statistical errors.

\begin{figure}
\epsfxsize=0.95\linewidth 
\centerline{\epsfbox{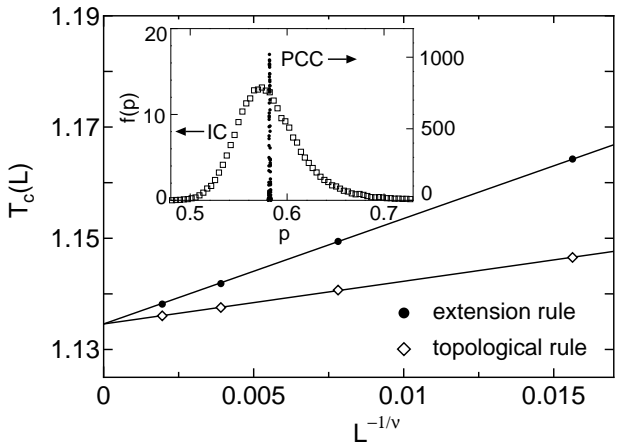}}
\vspace{2mm}
\caption{
Plot of $T_c(L)$ (in units of $J/k_B$) as a function of $L^{-1/\nu}$ 
for the $2d$ Ising model ($q=2$), where $\nu$=1. 
The system sizes are $L$ = 64, 128, 256, and 512. 
In the inset, the distribution of $p$, $f(p)$, for $L=64$ with 
the topological rule is shown for both the PCC and IC algorithms. 
Different scales are used for the vertical axis in the inset 
to express two quite different data. 
}
\label{fig1}
\end{figure}
Let us show the results of the $2d$ Ising model ($q=2$).  
In Fig.~\ref{fig1}, we plot the size dependence of $T_c(L)$ 
for both rules in units of $J/k_B$.  
We have determined $p_c(L)$ from the average of $p$, and 
calculated $T_c(L)$ through the relation 
$p_c(L) = 1 - e^{- J/ k_BT_c(L)}$. 
In this plot, as an illustration, we have used 
the known value of $\nu$ for the $2d$ Ising model ($\nu=1$).  
Using the least square method, we estimate $T_c$ as 1.1344(2) 
(1.1346(2)) 
for the extension (topological) rule, which is consistent 
with the exact value, $[\ln (1+\sqrt{2})]^{-1}$ = 1.1346. 
Here, the number in the parenthesis denotes the uncertainty in the 
last digit.
We have used the known value of $\nu$ but we may treat $\nu$ 
as an unknown parameter to be determined.  
Assuming the FSS relation, 
$ T_c(L) = T_c + a L^{-1/\nu}$, 
which is derived from Eq.~(\ref{scale}), 
we may follow the three-parameter ($T_c, 1/\nu, a$)
non-linear fitting procedure.  Then, we have obtained 
($T_c, 1/\nu$)=(1.1345(2), 1.00(4)) for the extension rule, 
and (1.1344(2), 1.04(4)) for the topological rule.  
Both estimates of $T_c$ and $\nu$ 
are consistent with the exact values.

We plot the distribution of $p$, $f(p)$, for $L=64$ with 
the topological rule, as an example, in the inset of 
Fig.~\ref{fig1}, which shows that the distribution of $p$ 
is sharply peaked at 0.58196 with the standard deviation 
of 0.0004 for our choice of the final $\Delta p$.  
For comparison, we also show $f(p)$ for the IC algorithm 
of the same size with the same rule; 
different scales are used for the vertical axis in the inset 
to express two quite different data. 
We notice that 
the distribution of $p$ for the IC algorithm is far broader.
A simple linear analysis yields that the width of 
the distribution for the PCC algorithm is proportional 
to $\sqrt{\Delta p/a}$, where $a$ is the value of $dE_p/dp$ 
at $p_c(L)$; using the FSS we expect $a \propto L^{1/\nu}$. 
It should be noted that we have obtained the expected 
$\Delta p$- and $L$-dependence for the width of 
the distribution of $p$. 
We use the average value of $p$ for the estimate of $p_c(L)$. 
Performing 10 runs, we have estimated $p_c(64)$ 
as $0.581954 \pm 0.000013$; the statistical errors 
are very small.  

\vspace{5mm}
\begin{figure}
\epsfxsize=0.95\linewidth 
\centerline{\epsfbox{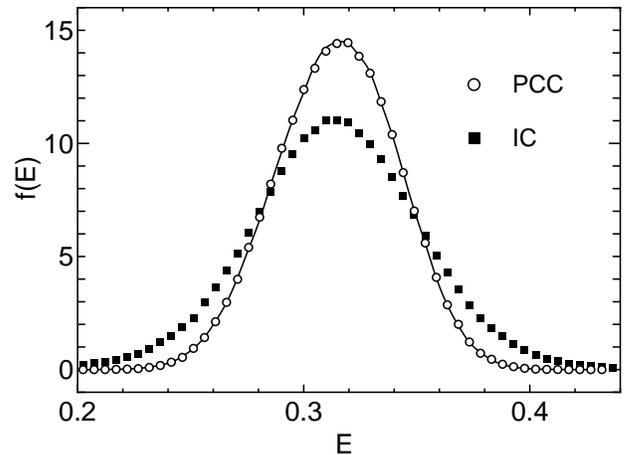}}
\vspace{2mm}
\caption{
The energy histogram, $f(E)$, of both the PCC and IC algorithms 
for $L=64$ with the topological rule for the 2d Ising model ($q=2$). 
The energy is measured in units of $J$. 
The energy histogram obtained by the constant-temperature 
calculation using the SW algorithm is also shown by a solid curve; 
the temperature has been chosen as 
$T_c(L)$ determined by the PCC algorithm.  
}
\label{fig2}
\end{figure}

\begin{figure}
\epsfxsize=0.95\linewidth 
\centerline{\epsfbox{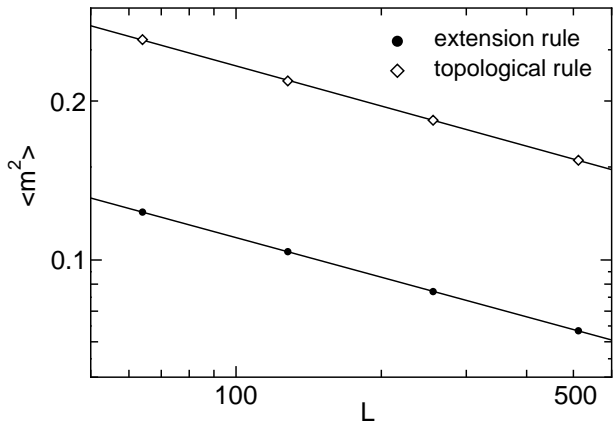}}
\vspace{2mm}
\caption{
Plot of $\l m^2 \r$ as a function of $L$ 
for the $2d$ Ising model ($q=2$) in logarithmic scale. 
}
\label{fig3}
\end{figure}
The resulting energy histogram, $f(E)$, of our PCC algorithm for 
$L=64$ with the topological rule is given in Fig.~\ref{fig2}. 
The energy histogram obtained by the constant-temperature 
calculation using the SW algorithm is shown 
by a solid curve in Fig.~\ref{fig2}; 
the temperature has been chosen as 
$T_c(L)$ determined by the PCC algorithm.  
The energy histogram of the PCC algorithm is indistinguishable 
from that by the constant-temperature calculation 
because of the sharp distribution of $p$.  
Thus, we may say that our ensemble is actually canonical 
for small enough $\Delta p$. 
The energy histogram of the IC algorithm, which is also 
given in Fig.~\ref{fig2}, has broad tails 
for both high-energy and low-energy sides. 
Although our ensemble is almost canonical, there are 
deviations in physical quantities, in principle;  
the variance of energy, $\l E^2 \r - \l E \r^2$, 
becomes larger with non-zero $\Delta p$, for example. 
We have checked the $\Delta p$-dependence of the systematic 
deviation for large $\Delta p$.  However, the deviation of 
the variance of energy is smaller than the statistical error, 
2 \% for $L=64$, with our choice of $\Delta p$. 

\begin{figure}
\epsfxsize=0.95\linewidth 
\centerline{\epsfbox{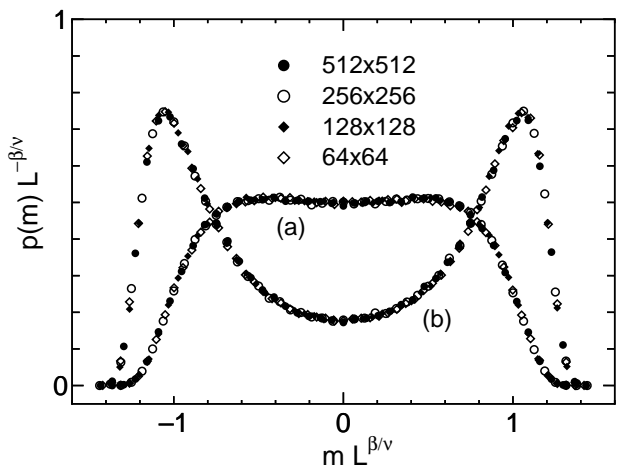}}
\vspace{2mm}
\caption{
FSS plot of $p(m)$ 
for the $2d$ Ising model ($q$=2), where $\beta/\nu$=1/8. 
The rules to determine percolating are the extension rule (a) 
and the topological rule (b).
}
\label{fig4}
\end{figure}
In order to estimate another critical exponent $\beta$, 
the magnetization exponent, we plot the average of the squared 
magnetization $\l m^2 \r$ as a function of $L$ 
in logarithmic scale in Fig.~\ref{fig3}. 
Since our Monte Carlo samples are sharply peaked at $p=p_c(L)$, 
in other words, at $T=T_c(L)$, we can use the FSS relation,
\begin{equation}
  \l m^2 \r_{\, T=T_c(L)} \sim L^{-2\beta/\nu},
\label{scale_m}
\end{equation}
for the estimate of $\beta/\nu$.  From the slopes of the data 
for both rules, we have $\beta/\nu$ = 0.125(2) (0.126(2)) 
for the extension (topological) rule, which is 
consistent with the exact value, 1/8 (=0.125).

It is quite interesting to study the distribution function 
of physical quantities.  We show the FSS plot of the distribution 
function $p(m)$ in Fig.~\ref{fig4}, based on the FSS relation,
\begin{equation}
  p(m)_{\, T=T_c(L)} \sim L^{\beta/\nu} f(mL^{\beta/\nu}) .
\label{scale_pm}
\end{equation}
The scaling plot for the extension rule (a) 
and that for the topological rule (b) are given there. 
The data for various sizes are collapsed on a single curve.  
We have very good FSS behavior for both rules.  
Two rules give different $t L^{1/\nu}$ in Eq.~(\ref{scale}) 
for $E_p=1/2$. 
It is easier to percolate for the extension rule compared 
with the topological rule. Therefore, $T_c(L)$ of the extension rule 
is higher than that of the topological rule, 
which results in the difference 
in the FSS functions for $p(m)$ between two rules. 

\begin{figure}
\epsfxsize=0.95\linewidth 
\centerline{\epsfbox{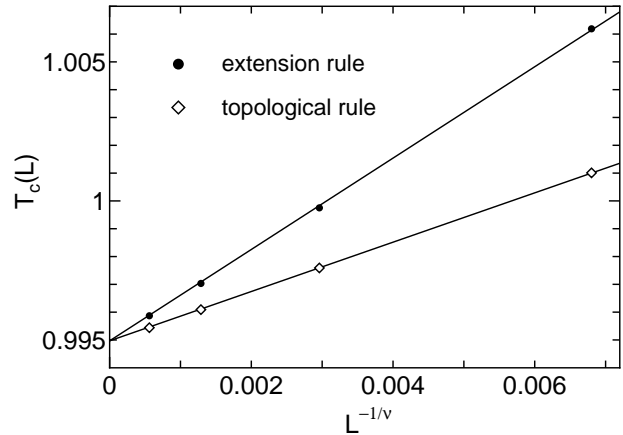}}
\vspace{2mm}
\caption{
Plot of $T_c(L)$ (in units of $J/k_B$) as a function of 
$L^{-1/\nu}$ for the $2d$ 3-state Potts model, where $\nu$=5/6. 
The system sizes are $L$ = 64, 128, 256, and 512. 
}
\label{fig5}
\end{figure}
Next turn to the case of the 3-state Potts model.  
The size dependence of $T_c(L)$ for both rules is shown 
in Fig.~\ref{fig5}. We have used the known value of 
$\nu$ for this plot; 
the exponent $\nu$ for the $2d$ 3-state Potts model 
conjectured by the conformal field theory \cite{Dotsenko84} is 5/6.  
We estimate the extrapolated value of $T_c$ as 
0.99490(6) (0.99494(6)) for the extension (topological) rule 
from Fig.~\ref{fig5}.  This value is consistent with the exact value, 
$[\ln (1+\sqrt{3})]^{-1}$ = 0.99497.  The convergence is very good 
for the 3-state Potts model. 
It is in contrast with the situation of the IC 
algorithm \cite{machta95}, where the convergence is 
not good enough maybe due to the problem of ``bottlenecks". 

We can estimate the critical exponent $\beta/\nu$ for the 
3-state Potts model from the size dependence of $\l m^2 \r$ 
as in the case of the Ising model. 
For the order parameter of the 3-state Potts model, 
we use the vector order parameter $(m_x,m_y)$.  
The $x$ and $y$ components of the vector order parameter 
are obtained from the three components, 
$m_1, m_2$ and $m_3$, as 
$ m_x = m_1 - (1/2)(m_2+m_3) $
and 
$ m_y = (\sqrt{3}/2)(m_2-m_3)$.
Using the FSS relation (\ref{scale_m}) 
for the 3-state Potts model, we have $\beta/\nu$ = 0.131(2) (0.134(2)) 
for the extension (topological) rule,  which is again consistent 
with the conjectured value, 2/15 (=0.1333) \cite{Dotsenko84}.  
We have also found nice FSS behavior for 
the order-parameter distribution functions $p(m)$ of
the 3-state Potts model as in the case of the Ising model. 
Here, $m$ stands for the absolute value of 
the vector order parameter.  
The details will be given in a separate paper.

Here we may remark on the computation time.  
We only need to modify 
the code of the SW algorithm slightly in order to change $p$.
The typical computation time to get $10^5$ Monte Carlo samples 
for $L=64$ is 471 seconds using the Alpha 21164A (533 MHz) machine, 
which is about 30 \% longer than that for the SW algorithm, 361 seconds. 
The most of time is spent on the usual procedure 
of the SW algorithm, that is, the assignment of the cluster 
and the cluster spin update.  
With a small cost of computation time, we can determine 
the critical point automatically without making simulations 
for various parameters.  In contrast, it takes 958 seconds 
in the case of the IC algorithm to get $10^5$ samples for $L=64$ 
because one should check whether the system is percolating 
several times before getting one Monte Carlo sample. 

To summarize, we have proposed a new cluster algorithm 
of tuning the critical point automatically.
Our algorithm is the extension of the SW algorithm \cite{sw87},
but we change the probability of connecting spins $p$ 
in the process of Monte Carlo spin update.  The resulting 
distribution of $p$ is sharply peaked at $p_c(L)$.  We approach 
the canonical ensemble in the limit of small $\Delta p$, 
which has been explicitly checked by the energy histogram. 
This is in contrast with the histogram of the IC 
algorithms \cite{machta95}, which has broad tails 
for both high-energy and low-energy sides. 
Thus, we can use the FSS analysis for the physical quantities. 
The estimated values of the critical temperatures and the critical 
exponents for the $2d$ Potts models are consistent with 
the exact values.
In order to get more accurate estimate of 
the critical point and critical exponents, the FSS analysis 
employed by Ferrenberg and Landau \cite{FL91} 
in a high-resolution Monte Carlo study is useful for the data 
obtained by the PCC algorithm; 
we extract $\nu$ first, and with $\nu$ determined 
quite accurately we can estimate $T_c$. 
Since the main purpose of 
the present Letter is to present a new and simple idea 
of the cluster algorithm, the refined data analysis 
including the corrections to FSS will be left to 
a subsequent study.

In the present study, we have shown the application of 
the PCC algorithm to the thermal phase 
transition of Potts models, but the idea is based only on the property 
of a percolation problem.  Thus, of course, we can use this algorithm 
in the study of the geometric percolation problem, that is, 
the $q=1$ Potts model, to get the percolation threshold, $p_c$, 
and various critical properties. 

Moreover, we can apply the PCC algorithm 
to any problem where the mapping to the cluster formalism exists. 
It is straightforward to apply this algorithm to the diluted 
Ising (Potts) model.  The lack of self-averaging has been recently 
discussed for the three-dimensional diluted Ising model 
\cite{AH96,WD98}.  The PCC algorithm is quite useful 
for the problem with the lack of self-averaging, where 
the distribution of $T_c$ due to the randomness is important, 
because we can tune the critical point of each 
random sample automatically. 
Another direction of application is the cluster quantum 
Monte Carlo simulation \cite{Evertz93,Kawashima94}, and this problem 
is left to a future study. 

We would like to thank N. Kawashima and M. Kikuchi 
for valuable discussions. 
This work was supported by a Grant-in-Aid for Scientific Research 
from the Ministry of Education, Science, Sports and Culture, Japan.


\end{multicols}
\end{document}